\documentclass[usenatbib]{mn2e}
\usepackage{upgreek}
\usepackage{amssymb,amsmath}
\usepackage{graphicx}
\usepackage{multirow}
\usepackage{natbib}
\usepackage{tabularx}
\usepackage{subfigure}
\def\cite{\citep}

\usepackage[usenames,dvipsnames]{color}

\title[Limits on FRB repeatability]{A survey of FRB fields: Limits on repeatability}
\makeatletter
\author[Petroff et al.]{E.~Petroff$^{1,2,3}$\thanks{Email: epetroff@astro.swin.edu.au}, S.~Johnston$^2$, E.~F.~Keane$^{4,1,3}$, W.~van Straten$^{1,3}$, M.~Bailes$^{1,3}$, \newauthor E.~D.~Barr$^{1,3}$, B.~R.~Barsdell$^{5}$, S.~Burke-Spolaor$^6$, M.~Caleb$^{7,1,3}$, D.~J.~Champion$^8$, \newauthor C.~Flynn$^{1,3}$, A.~Jameson$^{1,3}$, M.~Kramer$^{8,9}$, C.~Ng$^8$, A.~Possenti$^{10}$, B.~W.~Stappers$^{9}$
\\
$^1$Centre for Astrophysics and Supercomputing, Swinburne University of Technology, P.O. Box 218, Hawthorn, VIC 3122, Australia \\
$^2$CSIRO Astronomy \& Space Science, Australia Telescope National Facility, P.O. Box 76, Epping, NSW 1710, Australia \\
$^3$ARC Centre of Excellence for All-sky Astrophysics (CAASTRO) \\
$^4$SKA Organisation, Jodrell Bank Observatory, Cheshire, SK11 9DL, UK \\
$^5$Harvard-Smithsonian Center for Astrophysics, 60 Garden Street, Cambridge, Massachusetts, 02138, USA\\
$^6$Cahill Center for Astrophysics, California Institute of Technology, 1200 E California Blvd, Pasadena, CA 91125, USA\\
$^7$Research School of Astronomy and Astrophysics, Australian National University, ACT, 2611, Australia \\
$^8$Max Planck Institut f\"{u}r Radioastronomie, Auf dem H\"{u}gel 69, D-53121 Bonn, Germany \\
$^9$Jodrell Bank Centre for Astrophysics, University of Manchester, Alan Turing Building, Oxford Road, Manchester M13 9PL, United Kingdom \\
$^{10}$INAF - Osservatorio Astronomico di Cagliari, Via della Scienza 5, 09047 Selargius (CA), Italy \\
}
\makeatother

\date{}
\begin{document}

\maketitle

\begin{abstract}

Several theories exist to explain the source of the bright, millisecond duration pulses known as fast radio bursts (FRBs). If the progenitors of FRBs are non-cataclysmic, such as giant pulses from pulsars, pulsar-planet binaries, or magnetar flares, FRB emission may be seen to repeat. We have undertaken a survey of the fields of eight known FRBs from the High Time Resolution Universe survey to search for repeating pulses. Although no repeat pulses were detected the survey yielded the detection of a new FRB, described in \citet{PetroffFRB}. From our observations we rule out periodic repeating sources with periods $P$ $\leq$ 8.6 hours and rule out sources with periods 8.6 $< P <$ 21 hours at the 90\% confidence level. At $P$ $\geq$ 21 hours our limits fall off as $\sim$1/$P$. Dedicated and persistent observations of FRB source fields are needed to rule out repetition on longer timescales, a task well-suited to next generation wide-field transient detectors.

\end{abstract}

\begin{keywords}
methods: observational --- stars: magnetars --- pulsars: general --- radio continuum: stars
\end{keywords}

\section{Introduction}

Fast radio bursts (FRBs) are millisecond-duration radio pulses believed to be of extragalactic origin (see \citet{KeanePetroff} and references therein). Prior to the project on which this paper reports, almost all FRBs were found in high time resolution radio pulsar surveys that covered large swathes of sky in search of transient phenomena \cite{Lorimer07,Thornton13,Spitler14,SarahFRB}. The main objective of these surveys was to search for repeating transient phenomena such as pulsars and rotating radio transients (RRATs). The millisecond to sub-millisecond sampling times used for these surveys, combined with their large time on sky also made them excellent datasets in which to find FRBs.

The most successful of these surveys in terms of FRB yield has been the high Galactic latitude component of the High Time Resolution Universe survey (HTRU; \citeauthor*{Keith10} \citeyear{Keith10}) that found nine FRBs (\citeauthor{Thornton13}, \citeyear{Thornton13}; Champion et al., in prep.). While all bursts in this survey were found at high Galactic latitudes ($|b| > 15^\circ$) their distribution throughout the high latitude survey regions appears to be random. However no FRBs were found in the intermediate latitude portion of the survey \cite{Petroff14}. Since most of the FRBs were discovered years after occurrence no systematic follow-up was undertaken as part of the HTRU survey, which was completed in February 2014. This has changed with the advent of real-time detections such as that of FRB 131104, which was discovered in a targeted search of the Carina dwarf spheroidal galaxy and was observed repeatedly for a total of 78 hours in the year following detection with no further FRBs detected \cite{RaviShannon}.

Follow-up of FRBs on both short and long timescales is essential as it becomes increasingly important to solve the mystery of their origins. Although the true progenitors of FRBs are unknown, a cosmological origin is highly favoured \cite{Deng2014,Luan14,KeanePetroff,Katz2015}. Extragalactic theories currently under consideration include magnetar flares \cite{Thornton13,Kulkarni14}, blitzars \cite{Falcke}, supergiant pulses from neutron stars \cite{Cordes15}, and pulsar-planet systems \cite{Mottez2014}, all in galaxies at cosmological distances ($z > 0.2$). In the blitzar model an FRB is generated in a cataclysmic event and no repeat FRB emission is predicted. However, the magnetar flare, supergiant pulse, and pulsar-planet theories make specific predictions about FRBs as a repeating source on different timescales. 

Until recently it had been proposed that FRBs may share a common source with the terrestrial interference detected at Parkes known as `perytons' \cite{Burke11,Kulkarni14}. However the source of perytons has now been identified as on-site electronics that superficially appeared similar to FRBs \cite{Perytons2015}.

In the pulsar-planet model, the beamed radio emission is produced in the Alfv\'{e}n wings of a planet closely orbiting a pulsar. Such emission is constant but is only observable along our line of sight for $\sim$1 ms as the planet moves through its orbit. In such a case, the FRB would be a repeating event, occurring once per orbit. Ultra-light companion systems, such as those with a planetary companion, detected in our own Galaxy through pulsar timing have periods ranging from 1.56 hours to over 70 days, with a median period of 4 hours \cite{psrcat}. In such a scenario, recurring FRB events would be best observed through continuous monitoring of the field of a known FRB. No such emission was detected in the over 40 hours of follow-up conducted by \citet{Lorimer07} for FRB 010724 or in the extensive search by \citet{RaviShannon} for FRB 131104.

Energetic magnetar flares are more commonly detected through their X-ray emission and there are 28 known magnetars in our own Galaxy that have been found in X-ray searches including a radio-loud magnetar not associated with an X-ray outburst \cite{LevinMagnetar}. Of these 28 sources (of which only 4 are visible at radio frequencies) approximately 23 have been seen to burst and 3 have documented giant flares\footnote{From the online catalogue http://www.physics.mcgill.ca/~pulsar/magnetar/main.html} \cite{MagnetarCatalog}. If FRBs are produced in giant radio flares from a similar population in distant galaxies, long-term campaigns to re-observe the fields of known FRB fields over a period of several years or even several decades would be best-suited to detecting this kind of activity. A summary of proposed repeating progenitors and their timescales is presented in Table~\ref{tab:progenitors}.

\begin{table*}
\centering
\caption{Repeating progenitor models for FRBs and their timescales estimated by \citet{Mottez2014}, \citet{Turolla15}, and \citet{Cordes15} for pulsar-planet, magnetar giant flare, and supergiant pulse progenitors, respectively. Timescales for magnetar giant flares and supergiant pulsar pulses are given in terms of the number of events over the lifetime of a single source, and as the time between events if they are equally distributed throughout the progenitor's life.}\label{tab:progenitors}
\begin{tabular}{lll}
\hline \hline
Progenitor & Model & Timescale \\
\hline
Pulsar-planet & Highly periodic FRB from Alfv\'{e}n wings of a planet orbiting a pulsar & 1.5 hrs $ < P_\mathrm{orb} <$ 70 days  \\
 & & $P_\mathrm{orb,median}$ $\sim$ 4 hours \\
Magnetar giant flare & Pulse generated in the millisecond-duration giant flare of an  & $\sim$ 100 lifetime$^{-1}$ \\
 & extragalactic magnetar & $\sim$ 1 kyr \\
Supergiant pulsar pulses & Individual bright shot pulses from young, energetic neutron stars  & $\lesssim$ 10 lifetime$^{-1}$ \\
 & at cosmological distance & $\sim$ 1 Myr\\
\hline
\end{tabular}
\end{table*}

Discovery of repeat emission from an FRB source would definitively rule out some progenitor models, and the timescale of repetition could distinguish between others that explicitly predict repetition. Understanding the progenitors of FRBs would also lead to confirmation as to the distance to the source and whether or not FRBs are an extragalactic population. 

The first step in this process is to test predictions of FRB repeatability on a range of timescales by re-observing known FRBs to look for additional bursts. In this paper we describe a series of observations performed over 6 months at the Parkes radio telescope of 8 FRBs discovered in the HTRU survey (\citeauthor{Thornton13}, \citeyear{Thornton13}; Champion et al., in prep.). We describe our search strategy and data analysis and observations in Sections~\ref{sec:observations} and~\ref{sec:analysis}, the results of our survey in Section~\ref{sec:results}, and our limits on FRB repeatability in Section~\ref{sec:discussion}.

\section{Observations}\label{sec:observations}

The fields of eight known FRBs, listed in Table~\ref{tab:positions}, were surveyed over 110 hours between April and October 2014. The observations that were undertaken are listed in Table~\ref{tab:totalTime}. They were scheduled to allow for approximately monthly follow-up of each FRB field for 1$-$2 hours each. Each field was observed between 3 and 5 times in total over the 6 month period. The FRB positions were taken to be the coordinates of the beam centre from the discovery pointings, which are listed in Table~\ref{tab:positions}, however the true coordinates of each FRB are unknown due to the large uncertainty in the location of the source within the Parkes beam ($\sim$14$'$ beam width at full-width half-maximum at 1.4 GHz, \citeauthor{multibeam}, \citeyear{multibeam}). For this survey we performed observations in a 9$'$ offset grid around the beam centre positions as outlined in \citet{Morris2002} with 15-minute duration pointings. In this way we sampled the entire FRB field in each observing session. 

In the second observing session for this project a new FRB was discovered (FRB 140514) in a grid pointing 9$'$ offset from the field of FRB 110220 \cite{PetroffFRB}. Systematic follow-up of this source was absorbed into the survey with minimal gridding and longer total observing times per session. 

Additionally, we performed a focused search of the field of FRB 090625 for short-term repeatability by observing for 2.5 $-$ 8.6 hours per day over 5 days closely spaced. In these observations the pointing location was fixed at the beam centre position. FRB 090625 was observed on October 21, 23, 28, 29, and 30 for all available time while the source was above the horizon. The observation dates and total observing times are listed in Table~\ref{tab:totalTime}. 

\begin{table}
\centering
\caption{The Right Ascension and Declination of the detection beam centre position for all FRBs monitored in this survey, including the new discovery FRB 140514. The error radius for all surveyed FRBs is 7$'$, the full-width half-maximum for a single beam of the Parkes multi-beam receiver.  References are for [1] Champion et al., in prep., [2] \citet{Thornton13}, [3] \citet{DanThesis}, and [4] \citet{PetroffFRB}.}\label{tab:positions}
\begin{tabular}{cccc}
\hline \hline
FRB name & Right Ascension & Declination & Ref.\\
\hline
FRB 090625 & 03:07:47 & -29:55:35 & [1] \\
FRB 110220 & 22:34:38 & -12:33:44 & [2] \\
FRB 110626 & 21:03:43 & -44:44:19 & [2] \\
FRB 110703 & 23:30:51 & -02:52:24 & [2] \\
FRB 120127 & 23:15:06 & -18:25:37 & [2] \\
FRB 121002 & 18:14:47 & -85:11:53 & [3] \\
FRB 130626 & 16:27:06 & -07:27:48 & [1] \\
FRB 130628 & 09:03:02 & 03:26:16 & [1] \\
FRB 140514 & 22:34:06 & -12:18:46 & [4] \\
\end{tabular}
\end{table}

\begin{table*}
\centering
\caption{The total hours observed for each FRB in this campaign, including the additional source FRB 140514 which was discovered 2 months into the survey and was then observed in place of FRB 110220 for the remainder of the survey. The total time spent at the beam centre position from Table~\ref{tab:positions} and in gridding around the field are also listed.}\label{tab:totalTime}
\begin{tabular}{cccccc}
\hline \hline
FRB name & Observing date & Beam Centre & Gridding &  Total Observation duration & Total duration \\
& (UT) & (hours) & (hours) & (hours) & (hours) \\
\hline
FRB 090625 & 2014-05-14 & 0 & 0.75 & 0.75 & 33.65 \\
. & 2014-06-24/25 & 0.25 & 0.75 & 1 & \\
. & 2014-08-19 & 1.5 & 1 & 2.5 & \\
. & 2014-10-21 & 2.55 & 0 & 2.55 & \\
. & 2014-10-23 & 5.86 & 0 & 5.86 & \\
. & 2014-10-28 & 7.8 & 0 & 7.8 & \\
. & 2014-10-29 & 8.66 & 0 & 8.66 & \\
. & 2014-10-30 & 4.5 & 0 & 4.5 & \\
FRB 110220 & 2014-05-14 & 0.75 & 1 & 1.75 & 1.75 \\
FRB 110626 & 2014-04-21 & 0.5 & 2 & 2.5 & 11.25 \\
. & 2014-05-14 & 0.75 & 1 & 1.75 & \\
. & 2014-06-24 & 0.5 & 2 & 2.5 & \\
. & 2014-07-27 & 0.75 & 1 & 1.75 & \\
. & 2014-08-19 & 0.75 & 2 & 2.75 & \\
FRB 110703 & 2014-04-21 & 0.25 & 0.75 & 1 & 10.1 \\
. & 2014-05-14 & 0.75 & 1 & 1.75 & \\
. & 2014-07-27 & 1.5 & 1 & 2.5 & \\
. & 2014-08-19 & 1.5 & 1 & 2.5 & \\
. & 2014-09-30 & 0.5 & 1.25 & 1.75 & \\
. & 2014-10-30 & 0.6 & 0 & 0.6 & \\
FRB 120127 & 2014-05-14 & 0.75 & 1 & 1.75 & 5.5 \\
. & 2014-06-24 & 0 & 0.5 & 0.5 & \\
. & 2014-08-19 & 0.75 & 1 & 1.75 & \\
. & 2014-09-30 & 0.5 & 1 & 1.5 & \\
FRB 121002 & 2014-04-21 & 1 & 2 & 3 & 10.25 \\
. & 2014-05-14 & 0.75 & 1 & 1.75 & \\
. & 2014-06-24 & 0.75 & 1 & 1.75 & \\
. & 2014-07-27 & 1 & 1 & 2 & \\
. & 2014-08-19 & 0.75 & 1 & 1.75 & \\
FRB 130626 & 2014-04-21 & 1 & 2 & 3 & 9.5 \\
. & 2014-05-14 & 0.75 & 1 & 1.75 & \\
. & 2014-06-24 & 0.5 & 1 & 1.5 & \\
. & 2014-07-27 & 0.75 & 1 & 1.75 & \\
. & 2014-08-19 & 0.5 & 1 & 1.5 & \\
FRB 130628 & 2014-04-21 & 0.75 & 2 & 2.75 & 9 \\
. & 2014-05-14 & 0.75 & 1 & 1.75 & \\
. & 2014-06-24 & 0.25 & 0.75 & 1 & \\
. & 2014-07-27 & 1 & 1 & 2 & \\
. & 2014-08-19 & 0.5 & 1 & 1.5 & \\
FRB 140514 & 2014-05-14 & 1.2 & 0 & 1.2 & 19.2 \\
. & 2014-06-24 & 7.9 & 0 & 7.9 & \\
. & 2014-07-27 & 3.5 & 0 & 3.5 & \\
. & 2014-08-19 & 1 & 0 & 1 & \\
. & 2014-09-30 & 1.8 & 0 & 1.8 & \\
. & 2014-10-29 & 3.8 & 0 & 3.8 & \\
 \hline
 \end{tabular}
\end{table*}

\section{Data processing}\label{sec:analysis}

The observing mode for this survey is described in detail in \citet{PetroffFRB} where the new FRB discovery that resulted from this survey is presented. The observing system, based on the Berkeley Parkes Swinburne Recorder (BPSR) and the HI-Pulsar Signal Processor (HIPSR), incorporates two major upgrades that have become available since the HTRU survey, namely the real-time processing system and the ability to record 8-bit full-polarisation data from the linear feeds in the event of an FRB discovery. 

As outlined in \citet{PetroffFRB} the real-time transient pipeline on BPSR uses the \textsc{Heimdall} single pulse search software\footnote{http://sourceforge.net/projects/heimdall-astro/} to search for burst-like signals in the data while they are stored in the 120-s ring buffer on HIPSR. The data are searched over a range of pulse widths (0.128 $-$ 262 ms) and dedispersion trials (0 $-$ 2000 pc cm$^{-3}$) for candidates that satisfy the following criteria:

\begin{equation}\label{eq:livelimits}
\begin{gathered}
\mathrm{DM} \geq 1.5 \times \mathrm{DM}_\mathrm{MW} \\
\mathrm{S/N} \geq 10 \\
N_\mathrm{beams} \leq 4 \\
\Delta t \leq 8.192 \: \mathrm{ms} \\
N_\mathrm{events}(t_\mathrm{obs}-2\:\mathrm{s} \to t_\mathrm{obs}+2\:\mathrm{s}) \leq 5
\end{gathered}
\end{equation}

\noindent where DM is the dispersion measure, the electron column density along the line of sight, DM$_\mathrm{MW}$ is the total electron column density attributed to the Milky Way along that line of sight in the model by \citet{Cordes02}, S/N is the signal-to-noise ratio, N$_\mathrm{beams}$ is the number of adjacent beams in which the candidate occurs, $\Delta t$ is the pulse width, and $N_\mathrm{events}(t_\mathrm{obs}-2\:\mathrm{s} \to t_\mathrm{obs}+2\:\mathrm{s})$ is the total number of identified candidates within a 4 second window around the candidate, a final check to mitigate false positives due to radio frequency interference (RFI), which tends to occur as bursts closely spaced in time. 

Once an FRB matching the above criteria is identified, the 8-bit full-polarisation data around the time of the pulse is extracted from the buffer and saved to disk for all 13 beams. With this system it is now possible to record and recover the full-Stokes signal of a fast radio burst. The real-time burst search was performed for all data recorded during this survey at the time of observation. 

The real-time pipeline, which now operates on all data recorded with BPSR, is the first of two stages of processing to search the survey data for dispersed radio pulses. After the data are recorded at Parkes they are transfered to the Swinburne gSTAR supercomputing cluster via fibre link and stored on the supercomputer file system. The data are then processed again for potential pulses using a more thorough pipeline which does not run in real time. In this processing stage the data are searched using \textsc{Heimdall} from 0.128 $-$ 262 ms in pulse width, and from 0 $-$ 5000 pc cm$^{-3}$ over 9420 DM trials using a DM tolerance of 1.01 to avoid sensitivity loss due to poor trial spacing \cite{KeanePetroff}. A more thorough cleaning process is also performed to remove radio frequency interference in both frequency and time \cite{Kocz12}. Candidates satisfying the following criteria were flagged and inspected:

\begin{equation}\label{eq:frblimits}
\begin{gathered}
\mathrm{DM} \geq 5 \: \mathrm{pc} \: \mathrm{cm}^{-3} \\
\mathrm{S/N} \geq 8 \\
N_\mathrm{beams} \leq 4 \\
\Delta t \leq 8.192 \: \mathrm{ms} .
\end{gathered}
\end{equation}

Detection of a repeated FRB was defined as any single pulse identified in the field with a DM within 10\% of the original FRB detection. Variations in DM on several year timescales greater than 10\% of the total value would only occur if a large fraction of the ionised material was local to the progenitor as such variations are orders of magnitude greater than those observed from interstellar turbulence \cite{Keith13,PetroffDM}. Physical constraints on dense environments around the FRB progenitor have been made by \citet{Luan14} making such large variations in DM for FRB progenitors unlikely. However, the full range of Galactic and extragalactic DMs were searched to look not only for repeated FRB pulses, but also for any yet undiscovered pulsars or rotating radio transients that may lie in the survey fields. 

\section{Results}\label{sec:results}

Only one significant dispersed pulse was detected in the 110 hours of observations in the real-time pipeline: the new FRB 140514 \cite{PetroffFRB}. This burst was detected at a 9 arcmin offset from the beam centre position of FRB 110220 in beam 1 of the multibeam receiver and was correctly identified in the real-time pipeline with a S/N of 16. The DM of FRB 140514 was found to be 562.7(6) pc cm$^{-3}$ while that of FRB 110220 was 944.38(5) pc cm${^-3}$ \cite{Thornton13}. All other events could be classified as radiometer noise or RFI.


Due to the difference in DM the bursts were judged to be separate events. However, this conclusion assumes that a single progenitor cannot produce two bursts of different DM separated in time by several years. It is highly unlikely that the bulk of the line-of-sight electron column density has changed so substantially \cite{Luan14,Artem14}. It remains the case that the progenitor itself could be enshrouded in ionised material which could vary significantly in density over time causing a large DM change. This would then place the source at a much smaller distance, perhaps in the local Universe as has been proposed by \citet{Pen2015} and \citet{Connor2015} and with much lower total energies. 

The discovery of FRB 140514 offered an unprecedented opportunity for immediate and sustained follow-up of an FRB field during the weeks and months after the event.The field of the new FRB was observed during all subsequent observing sessions including an 8-hour track on 24 June 2014 for the entire time the field was observable from Parkes. In total the field of FRB 140514 was observed for 19.2 hours in the 5 months after the observed radio burst \cite{PetroffFRB}. 

The 120 hours of Parkes data were also searched using the deeper search pipeline described in Section~\ref{sec:analysis} for pulses occurring in the data at any DM above the zero-DM RFI threshold (DM $>$ 5 pc cm$^{-3}$). The deeper pipeline found no additional dedipsersed pulses down to an S/N of 8 in the full data set including pulses within the range of Galactic dispersion measures that might be attributed to pulsars or RRATs. 

\section{Discussion}\label{sec:discussion}

In the following subsections we will discuss the implications of the non-detections on FRB progenitor models in the context of total observing time, the multi-day monitoring of FRB 090625, and the prompt follow-up in the months following FRB 140514.

\subsection{Total time}

The total observing time of 110 hours spaced roughly monthly over a 6-month period is insufficient to place substantial limits on infrequently occurring flares from bursting sources such as magnetars or supergiant pulses from extragalactic neutron stars, as noted in \citet{Cordes15}. Our strongest limits on repetition from a single source during our observations comes from FRB 090625 which was observed for a total of 33.65 hours with no detected pulsed emission. 

To place stronger constraints on these types of events would require hundreds of hours of monitoring over multiple years. The anticipated timescale between magnetar giant flares (years to decades) \cite{MagnetarCatalog} and repetition timescales of 1000s of years for supergiant pulses \cite{Cordes15} makes the probability of catching repeats extremely low. Ultimately stronger limits on long-term repeatability will come from wide-field radio telescopes capable of monitoring these fields as part of routine sky surveys. Dedicated time on telescopes with small field of view, such as Parkes, is difficult to justify given the amount of time needed. Systematic follow-up of future FRB discoveries made in future surveys and with future instruments will also be necessary to monitor these source fields in the months and years after a detection.

\subsection{Multi-day observations of FRB 090625}

\citet{Mottez2014} have predicted that a planet within the pulsar wind could produce a strictly periodic FRB signal that repeats as the period of the planetary orbit. To place constraints on repetition on short timescales, we undertook a multi-day observing campaign for a single FRB. FRB 090625 was chosen for this additional observing as it was above the horizon for all time slots available to the project. The observations were conducted over five nights: 2014 October 21, 23, and 28$-$30. Total time spent on the source was 29.4 hours, and two exceptionally long tracks of 7.8 and 8.6 hours were achieved on October 28 and 29, respectively, see Table~\ref{tab:090625}. 

\begin{table}
\centering
\caption{Summary of observations for a multi-day campaign in the field of FRB 090625}\label{tab:090625}
\begin{tabular}{cc}
\hline \hline
UTC start & T$_{\mathrm{obs}}$ (hours) \\
\hline
2014-10-21 14:49:34 & 2.55 \\
2014-10-23 10:49:04 & 5.86 \\
2014-10-28 10:52:46 & 7.8 \\
2014-10-29 10:06:00 & 8.6 \\
2014-10-30 10:53:53 & 4.5 \\
\end{tabular}
\end{table}

\begin{figure}
\centering
\includegraphics[width=8cm]{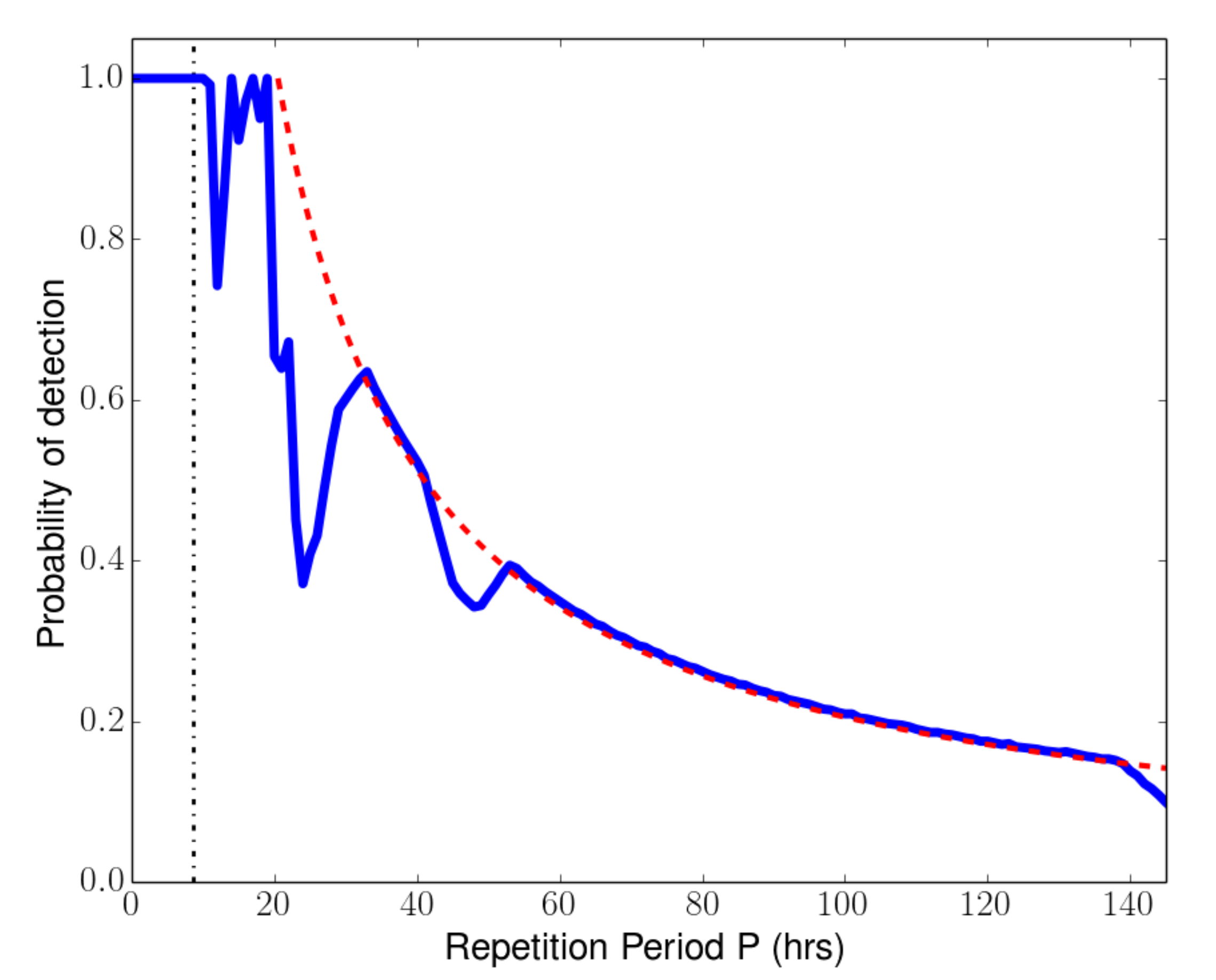}
\caption{Probability of detection for repeating progenitors with a repetition period $P$ in the 5 day campaign for FRB 090625. Sources with periods less than our longest observation ($P <$ 8.6 hrs, dot-dashed line) are ruled out. Periods $P <$ 21 hours are also ruled out with 90\% confidence. At $P$ = 21 hours the probability of detection drops off as 1/$P$ (red dashed line). This limit assumes that FRBs are strictly periodic. \label{fig:periods}}
\end{figure}

With these observations we can rule out a repeating progenitor system with a period ($P$) of less than 8.6 hours, the longest continuous observation in the campaign assuming that any repeat emission is above the flux limit of a Parkes beam, $S \gtrsim 0.5$ Jy. Due to the spacing of our observations we can also rule out repeating progenitors with periods 8.6 $< P <$ 21 hours with 90\% confidence. For $P >$ 21 our probability of detecting repeat emission, assuming the source emits a pulse on every rotation, decreases as 1/$P$ with the exception of some poor sensitivity to certain periods due to observation spacing, Figure~\ref{fig:periods}.

Limits on a periodic repeating progenitor can be similarly placed for each source monitored in this campaign. The longest continuous observation $t_\mathrm{obs,max}$ of a single source places a hard limit on repetition periods $P \leq t_\mathrm{obs,max}$ and a 90\% confidence limit on periods $P \lesssim 2 \times t_\mathrm{obs,max}$ after which sensitivity decreases as approximately 1/$P$, as in the case of FRB090625. 

\subsection{Follow-up of FRB 140514} 

Before the advent of real-time transient detection it was not possible to monitor the field of an FRB in the days and weeks after it occurred for pulses that might be associated with an active period of flaring or relaxation to a rest-state. Such pulses would give valuable clues about the events producing the observed bursts. The immediate discoveries of FRB 131104 \cite{RaviShannon} and FRB 140514 \cite{PetroffFRB} enabled rapid follow-up on a timescale never before available. The longest observation of FRB 140514 conducted in this survey was undertaken 41 days after the event and consisted of a continuous 7.9 hour observation at the position of the discovery beam. Observations of the field were also performed 7 hours, 41 days, 74 days, 97 days, 138 days, and 168 days after the event in which no repeat emission was detected. 

The probability of detecting the new FRB in our observations based on the total time on sky, and given the \citet{Thornton13} rate for an isotropic distribution of sources is 33.5\%. The revised, lower rate from Champion et al. (in prep.) based on a full search of the HTRU high latitude survey gives a probability of a new FRB detection in our survey of 25.7\%, with a substantially higher probability (68\%) of detecting no new bursts. Even with the lower rate, the probability of detecting a new event is still not negligible and we conclude that the detection of FRB 140514 in our survey is not entirely unexpected.

\section{Conclusions}

We present the results of a survey of known FRB fields to place limits on FRB repeatability. The total survey consisted of 110 hours over 6 months dedicated to re-observing the fields of 8 known sources. No repeat emission was detected from an FRB during this time placing weak limits on bursting or flaring sources; a more detailed and long-term study would be needed to rule out progenitors such as magnetar flares or supergiant pulses from extragalactic neutron stars. One component of this survey consisted of a multi-day campaign to observe a single FRB field and place limits on short-term repetition. From this sub-study we rule out repeating progenitors with periods less than 8.6 hours and place limits on repetition for periods between 8.6 and 21 hours at the 90\% confidence level. We are also able to constrain systems with greater orbital periods making pulsar-planet systems unlikely progenitors for FRBs.

In the course of this survey a new FRB was detected near the field of FRB 110220 and determined to be independent from the previous source due to difference in DM. Further effort is required to refine limits on repetition of FRB sources. A dedicated monitoring campaign is not feasible using single dish telescopes (like Parkes, even with a phased array feed) with a small field of view, and instead might be better-suited to wide field interferometric telescopes with high time resolution observing capabilities, such as UTMOST\footnote{http://astronomy.swin.edu.au/research/utmost}, MeerKAT \cite{Obrocka2015}, or SKA1 \cite{SKATransients}. Re-observation of an FRB in the days after a detection could provide valuable information about potential periods of high activity or relaxation experienced by the progenitor and would yield further insight into the origin of these bursts. Real-time detections of FRBs with future surveys should then be systematically followed up to search for such emission.

\section*{Acknowledgments}

We thank the referee for the useful comments and suggestions related to this manuscript. The Parkes radio telescope and the Australia Telescope Compact Array are part of the Australia Telescope National Facility which is funded by the Commonwealth of Australia for operation as a National Facility managed by CSIRO. Parts of this research were conducted by the Australian Research Council Centre of Excellence for All-sky Astrophysics (CAASTRO), through project number CE110001020. This work was performed on the gSTAR national facility at Swinburne University of Technology. gSTAR is funded by Swinburne and the Australian Government’s Education Investment Fund. EP would like to thank P. Edwards for assisting with scheduling additional observations for this work.

\bibliographystyle{mnras}
\bibliography{journals,P871} 

\end{document}